\documentclass[preprint,11pt]{elsarticle}

\usepackage{amssymb}
\journal{Nuclear Physics B}

\begin{document}

\begin{frontmatter}



\title{Climate Stability of Habitable Earth-like Planets}


\author[km,km2]{Kristen Menou}

\address[km]{Centre for Planetary Sciences,
    Department of Physical \& Environmental Sciences, University of
    Toronto at Scarborough, Toronto, Ontario M1C 1A4,
    Canada}
\address[km2]{Department of Astronomy \& Astrophysics,
    University of Toronto, Toronto, Ontario M5S 3H4, Canada}

\begin{abstract}
The carbon-silicate cycle regulates the atmospheric $CO_2$ content of
terrestrial planets on geological timescales through a balance between
the rates of $CO_2$ volcanic outgassing and planetary intake from rock
weathering. It is thought to act as an efficient climatic thermostat
on Earth and, by extension, on other habitable planets. If, however,
the weathering rate increases with the atmospheric $CO_2$ content, as
expected on planets lacking land vascular plants, the carbon-silicate
cycle feedback can become severely limited. Here we show that
Earth-like planets receiving less sunlight than current Earth may no
longer possess a stable warm climate but instead repeatedly cycle
between unstable glaciated and deglaciated climatic states. This has
implications for the search for life on exoplanets in the habitable
zone of nearby stars.
\end{abstract}

%\begin{keyword}



%\end{keyword}

\end{frontmatter}


\section{Introduction}

It is generally thought that the carbon-silicate cycle acts as a
stabilizing feedback and a powerful thermostat for the Earth climate,
guaranteeing surface liquid water conditions. Above freezing
temperatures, rock weathering occurs faster at higher temperatures,
which reduces the $CO_2$ atmospheric partial pressure, $pCO_2$, and
cools down the climate. Conversely, on a frozen planet that
temporarily lacks weathering, atmospheric $CO_2$ builds up from
continued volcanic outgassing, which warms up the climate until
surface liquid water and weathering conditions are restored
\cite{1,2,3}. Such $pCO_2$ build up is in fact the leading scenario
for the deglaciation of Earth following a snowball event
\cite{4,5,6}. Generalizations of these concepts to Earth-like planets
around other stars are central to the definition of their liquid water
habitable zone \cite{2,7,8}. In particular, planets subject to modest
levels of insolation are expected to achieve temperate conditions with
liquid water at the surface by building up massive enough $CO_2$
atmospheres \cite{2,9,10}.

Experimental data, theoretical arguments and paleoclimate modeling
suggest that the rate of $CO_2$ intake via rock weathering by a planet
lacking land vascular plants increases with the atmospheric $CO_2$
content, $pCO_2$ \cite{3,11,12,13,14}.  This feature of lifeless
planets or planets with only primitive forms of life is important
because it will limit the buildup of $CO_2$ at high values. As a
result, the climate thermostat due to the carbon-silicate cycle should
become less efficient on weakly-insolated Earth-like planets located
in the outer regions of the habitable zone.

\section{Climate-Weathering Models}

To address this issue quantitatively, we model Earth-like climates
with a zero-dimensional, energy balance model that equates the net
insolation and thermal radiation fluxes,
\begin{equation}
\frac{S}{4} \left[ 1-\alpha(T_{\rm surf},pCO_2) \right]= OLR(T_{\rm surf},pCO_2),
\end{equation}
where $S$ is the insolation flux, $\alpha$ is the planetary Bond
albedo and $OLR$ is the outgoing longwave radiation flux emitted by
the planet. $OLR$ and $\alpha$ are functions of the surface
temperature, $T_{\rm surf}$, and $pCO_2$, derived from
radiative-convective climate models \cite{9} (see Appendix~A for
details).

While we assume that the climate reaches thermal equilibrium rapidly,
by virtue of Equation~(1), the slower CO$_2$ compositional
equilibrium is not imposed a priori in our models. Rather, $pCO_2$ is
evolved on the relevant geological timescales according to
\begin{equation}
\frac{d}{dt} pCO_2= V-W(T_{\rm surf},pCO_2),
\end{equation}
where $V$ is the global $CO_2$ volcanic outgassing rate (estimated as
$V_\oplus =7$~bars/Gyr for Earth \cite{16}) and $W$ is the rate of
$CO_2$ intake by the solid planet via rock weathering.  The functional
form of $W$ is adapted from Earth studies for pre-vascular plant
conditions \cite{14}:
\begin{equation}
\frac{W}{W_\oplus} = \left(\frac{pCO_2}{p_\oplus}\right)^\beta
e^{\left[ k_{\rm act} (T_{\rm surf} -288)\right]} \times \left[ 1+
  k_{\rm run} (T_{\rm surf} -288)\right]^{0.65},
\end{equation}
where $p_\oplus=330 \, \mu$bar is the pre-industrial $pCO_2$ level,
$W=W_\oplus \equiv V_\oplus$ for $T_{\rm surf}=288$K, $k_{\rm
  act}=0.09$ is related to an activation energy and $k_{\rm
  run}=0.045$ is a runoff efficiency factor. Varying \cite{14} $k_{\rm
  act}$ in the range $0.06$--$0.135$ and $k_{\rm run}$ in the range
$0.025$--$0.045$ has only a minor quantitative impact on our results.
Values of $\beta=0.25$--$1$ have been considered in the literature
\cite{3, 12, 19, 21} for the dependence of weathering on $pCO_2$ in
the absence of land vascular plants. We use $\beta=0.5$ (default) and
$0.25$ in this work. Note that equations~(1) and~(2) are coupled
through $pCO_2 $ and $T_{\rm surf}$.

\section{Climate Solutions}

\subsection{Steady-State Solutions}

Steady-state climate solutions, satisfying both radiative
(Equation~(1)) and weathering equilibrium ($d /dt \equiv 0$ in
Equation~(2)) are represented in Figure~1 by intersecting cooling and
heating curves.  Solid red lines in Figure~1 show the albedo-corrected
insolation flux (LHS of Equation~(1)) received by a planet located at
1, 1.25 and 1.6 AU from a Sun-like star as a function of surface
temperature.\footnote{Even though the planetary albedo $\alpha$ in
  Equation (1) depends on $pCO_2$, we find that this dependence is
  quantitatively negligible for $pCO_2 \ll 0.2$~bar. For simplicity,
  we plot heating (red) curves in this low $pCO_2$ limit in Figure~1,
  which is indeed satisfied by all the cooling (blue) curves
  shown. All our other results fully account for the
  $\alpha$--$pCO_2$ dependence.} Net insolation drops precipitously
from 273K to 263K as the planetary surface freezes and the albedo
reaches $\sim 0.65$.  The various blue lines in Figure~1 represent the
$OLR$ cooling flux (RHS of Equation~(1)) according to various
scenarios for the atmospheric $CO_2$ content.

A standard model with $pCO_2$ arbitrarily fixed at $p_\oplus=330 \,
\mu$bar (i.e. not constrained by Equation~(2)) is represented by the
slanted solid blue line. As is well known \cite{22,23,24}, three
steady-state climate solutions exist in such a model for an Earth-like
planet at 1AU, as indicated by diamonds where the red and blue curves
intersect. Earth's current climate ($T_{\rm surf} \simeq 288$~K) and a
globally-frozen state ($T_{\rm surf} \simeq 229$~K) are both stable,
while the intermediate state ($T_{\rm surf} \simeq 270$~K) is
thermally unstable.

However, when $pCO_2$ is also required to satisfy weathering
equilibrium (Equation~(2)), steady-state climate+weathering solutions
only exist above freezing temperatures since weathering stops
operating on a frozen planet. A model including weathering without any
$pCO_2$ dependence\cite{9} ($\beta =0$ in Equation~(3)) is represented
by the vertical solid blue line in Figure~1. In such a model,
weathering equilibrium enforces a unique surface temperature, set by
requiring that weathering balances volcanic outgassing in
Equation~(2), and $pCO_2$ is only indirectly constrained by the
constant $T_{\rm surf}$ requirement.

On the other hand, when the weathering rate depends on both $pCO_2$
and $T_{\rm surf}$, noticeable bends appear in the blue cooling curves
shown in Figure~1. Indeed, in this class of models, the reduced
efficiency of weathering at low temperatures must be balanced by large
$pCO_2$ values to match the volcanic outgassing rate $V$. The weaker
the weathering $pCO_2$ dependence, the stronger is the $pCO_2$
build-up at low surface temperatures (compare $\beta=0.5$ and $0.25$
models represented by the dashed and dash-dotted lines in Figure~1,
respectively). A planet with a larger volcanic $CO_2$ outgassing rate
\cite{25} achieves a warmer stable climate ($T_{\rm surf} \simeq
292$~K for $V=3 V_\oplus$ and adopting our default weathering
parameters, which is the case shown as a dotted line in Figure~1).

Interestingly, stable climate+weathering solutions can cease to exist
at low insolation levels, such as the 1.25 and 1.6 AU cases shown in
Figure~1, for a strong enough dependence of the weathering rate on
$pCO_2$. For example, we find that blue curves no longer intersect
with heating (insolation) lines beyond 1.077 AU if $\beta =0.5 $ and
beyond 1.25 AU if $\beta =0.25 $. Conversely, a very weak weathering
dependence on $pCO_2$ ($ \beta < 0.1$), including the singular case
$\beta =0$ (vertical solid blue line in Figure~1), do permit stable
climate+weathering solutions at almost arbitrarily low insolation
levels \cite{9}. Low $\beta$ values may be the relevant limit for
planets where land vascular plants are widespread
\cite{3,11,12,13,14,21}. On the other hand, for values of $\beta \sim
0.25$--$0.5$ appropriate for planets lacking land vascular plants
\cite{3,11,12,13,14,21}, the bending of cooling curves seen in
Figure~1 also implies that at fixed volcanism rate, $V$, less
insolated planets achieve climate+weathering equilibrium at gradually
lower $T_{\rm surf}$ values. For example, in the case $\beta =0.25$ ,
equilibrium is only marginally achieved above freezing temperatures at
1.25 AU, as shown in Figure~1. At low enough insolation levels (far
enough away from the star), climate+weathering equilibrium is no
longer possible above freezing temperatures, which implies that
weathering equilibrium is unattainable.

On planets lacking weathering equilibrium, the climate must repeatedly
cycle through a succession of radiative equilibria as illustrated in
Figure~1: rapid transition from marginal to full glaciation
(A$\rightarrow$B, at fixed $pCO_2$), slow build-up of $pCO_2$ caused
by volcanic outgassing in the absence of weathering (B$\rightarrow$C),
rapid transition to a deglaciated state (C$\rightarrow$D, at fixed
$pCO_2$) and gradual $pCO_2$ decay under the action of weathering
(D$\rightarrow$A), until the cycle repeats again with full
glaciation. The general properties of the four critical points A-B-C-D
of this climate cycle, which are independent of details of the
weathering model, are quantified in Figure~2 as a function of orbital
distance from a Sun-like star. Blue curves correspond to the coldest
cycle point with the lowest $pCO_2$ value, B, while the red curves
correspond to the hottest point with the highest $pCO_2$ value, D.

Figure~2 shows that Earth-like planets at larger orbital distances
glaciate and deglaciate at larger $pCO_2$ values. A deglaciation with
$pCO_2 \simeq 0.14$~bar at $1$~AU is consistent with values reported
for snowball Earth deglaciation \cite{6}. Planets beyond $1.3$~AU
support massive ($>0.3$~bar) $CO_2$ atmospheres throughout their
climate cycle. Planets in the frozen state have albedos $\simeq 0.65
$, while the albedo of unfrozen planets rises from $\simeq 0.3$ to
$0.45$ in the range $1$-$1.8$~AU, from an increasing atmospheric
scattering contribution. The extremes of surface temperature along the
cycle vary modestly with insolation level, with $T_{\rm surf} \simeq
210$-$225$~K at the coldest point and $310$-$330$~K at the hottest
point.

\subsection{Climate Cycles}

Explicit time-dependent integrations of the system of
Equations~(1)-(3) reveal details of the climate cycle illustrated in
Figure~1. We initiated these integrations at the hot, high $pCO_2$
(weathering-independent) point D and confirmed that Earth-like planets
receiving sufficiently large insolation fluxes settle to a
steady-state warm climate solution after relaxation to weathering
equilibrium. By contrast, planets at large enough orbital distances
(low enough insolation levels) experience large amplitude climate
cycles, as anticipated from our discussion of equilibrium solutions in
relation to Figure~1.

Figure~3 shows six illustrative examples (A-F) of such climate cycles,
shown in terms of variable $T_{\rm surf}$ and $pCO_2$ curves. Most of
the cycle time is spent in the frozen state, during which $pCO_2$
build-up is slow compared to the fast weathering that occurs at
above-freezing temperatures. In model 1, 4.6\% of the 70 Myr cycle is
spent in a warm state with surface liquid water. The corresponding
numbers are 0.8\% of 477 Myr in model 2, 7.5\% of 76 Myr in model 3,
3.9\% of 139 Myr in model 4, 28\% of 312 Myr in model 5 and 17\% of
517 Myr in model 6. Faster weathering at higher $T_{\rm surf}$ also
implies that most of the time in the unfrozen state is spent just
above freezing temperatures, near the lowest $pCO_2$ levels covered
during the cycle. For a fixed volcanic outgassing rate, $V$, the
climate cycle duration increases with decreasing insolation because
larger absolute $pCO_2$ values must be reached for climate transitions
to occur (Figure~2). Decreasing insolation also reduces the fraction
of cycle time spent with surface liquid water by the planet, although
this can be compensated for by stronger volcanic outgassing. A weaker
$pCO_2$ weathering dependence (lower $\beta$) lengthens the duration
of the unfrozen state since the decline in $pCO_2$ with time has less
of an effect on the weathering rate.

To summarize, a temperature-only dependence of the weathering rate
($\beta =0$) uniquely ties the surface temperature to the volcanic
outgassing rate $V$ via Equation~(2). A more general dependence on
$pCO_2$ ($\beta >0$) leads to a richer set of climate solutions,
including unstable climate cycles at low enough insolation levels,
when weathering equilibrium ceases to exist. These results are not
specific to the weathering functional form adopted in Equation~(3), in
the sense that other weathering laws with a positive dependence on
$pCO_2$ and $T_{\rm surf}$ would lead to qualitatively similar climate
behaviors.

\section{Conclusions}

The key new feature of our analysis is the lack of stable climates on
Earth-like planets lacking land vascular plants, at low enough
insolation levels. This suggests that a subset of Earth-like planets
located in the outer regions of habitable zones may be preferentially
found in a frozen, rather than deglaciated, state. A globally frozen
state might be observationally inferred from the very high albedo and
the correspondingly low water content of the planet's
atmosphere. According to these results, some Earth-like planets in the
outer habitable zone would also be caught in a transiently warm state
with surface liquid water present only infrequently.

The link between unstable climate cycles and the emergence and
evolution of life on weakly-insolated Earth-like planets is unclear
but possibly important. A reduced amount of time with surface liquid
water on planets experiencing climate cycles could in principle slow
down the emergence and/or evolution of life. On the other hand, life
itself could strongly impact the weathering process on
weakly-insolated Earth-like planets, as it seems to have done on early
Earth \cite{3,11,12,13,14,21}. In particular, the ability of land
vascular plants to regulate the soil $pCO_2$ level that is relevant to
the weathering process, well above atmospheric $pCO_2$ levels, is
consistent with these plants effectively decoupling the weathering
rate from the atmospheric $pCO_2$ level \cite{3,11,12,13,14,21},
leading to $\beta \rightarrow 0$ in Equation~(3). As a result
(Figure~1, vertical line), the climate of weakly-insolated Earth-like
planets could be stabilized against transient cycles once the presence
of land vascular plants becomes widespread.  This would constitute a
strong example of life exerting a feedback on its environment.

It is worth noting that Earth's geological record is qualitatively
consistent with the evolutionary path one may envision for a habitable
planet orbiting a star that is gradually brightening over
time. Repeated snowball events should be restricted to early times,
when insolation is weak and land vascular plants are absent. They
should disappear at late times once insolation is strong enough and/or
land vascular plants become widespread.

\section*{Acknowledgments}

This work was supported by the Natural Sciences and Engineering
Research Council of Canada. The author is grateful to J. Leconte and
D. Valencia for comments on the manuscript.

\begin{figure}
\centerline{\includegraphics[width=1.\textwidth]{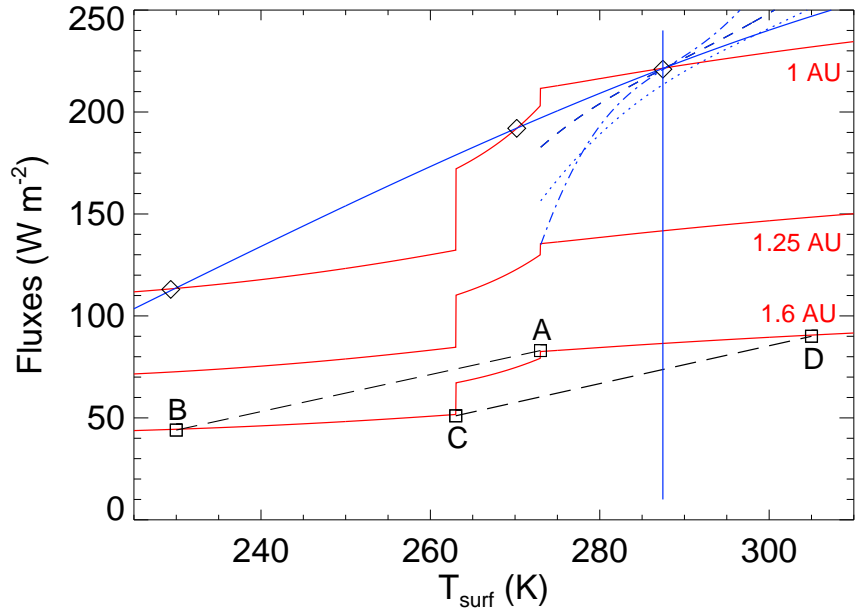}}
\caption{Climate at global radiative equilibrium for an Earth-like
  planet. Red lines show the albedo-corrected insolation (heating)
  flux as a function of surface temperature, $T_{\rm surf}$, at 1~AU
  (top), 1.25~AU (middle) and 1.6~AU (bottom) from a Sun-like
  star. Blue lines show the infrared cooling flux ($OLR$) according to
  various scenarios for the atmospheric $CO_2$ content (slanted solid
  line: fixed $pCO_2$ model; dashed: $\beta=0.5$ weathering model;
  dotted: $\beta=0.5$ weathering model with 3 times larger $CO_2$
  outgassing rate; dashed-dotted: $\beta=0.25$ weathering model;
  vertical solid: $\beta=0$ weathering model).  When no stable climate
  exists at large orbital distances (absent blue-red intersections),
  the climate must repeatedly cycle through points A-B-C-D shown in
  the 1.6~AU case, with a slow $pCO_2$ build-up (B-C), a transition to
  a hot climate (C-D), a weathering period with decreasing $pCO_2$
  (D-A) and a transition to global glaciation (A-B).}\label{afoto}
\end{figure}

\begin{figure}
\begin{center}
\centerline{\includegraphics[width=1.\textwidth]{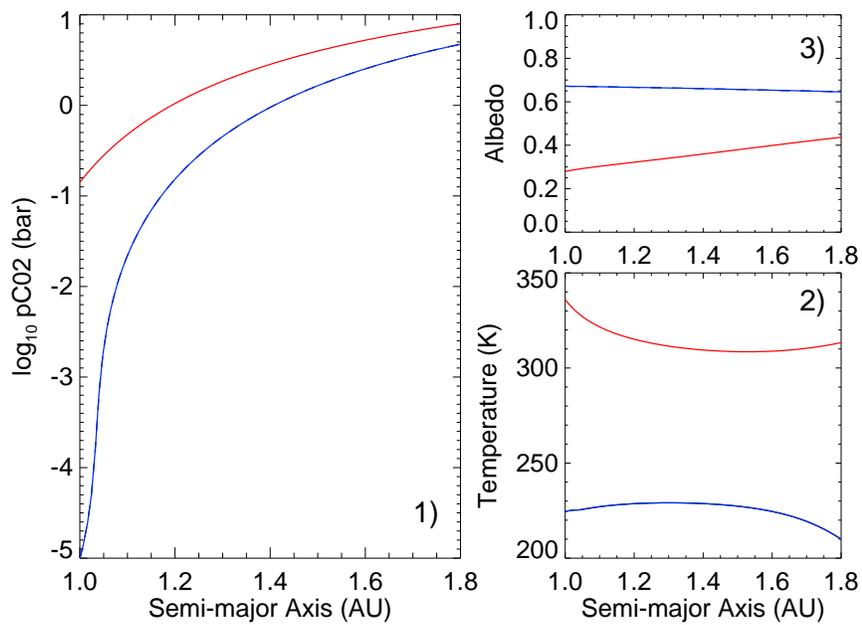}}
\caption{Values of atmospheric partial $CO_2$ pressure (1), surface
  temperature (2) and planetary albedo (3) at extremes of the climate
  cycle illustrated in Figure~1, as a function of orbital distance
  from a Sun-like star. Blue curves correspond to point B (cold, high
  albedo, low $pCO_2$) and red curves to point D (hot, low albedo,
  high $pCO_2$) of the cycle shown in Figure~1.}\label{afoto2}
\end{center}
\end{figure}

\begin{figure}
\begin{center}
\centerline{\includegraphics[width=.55\textwidth]{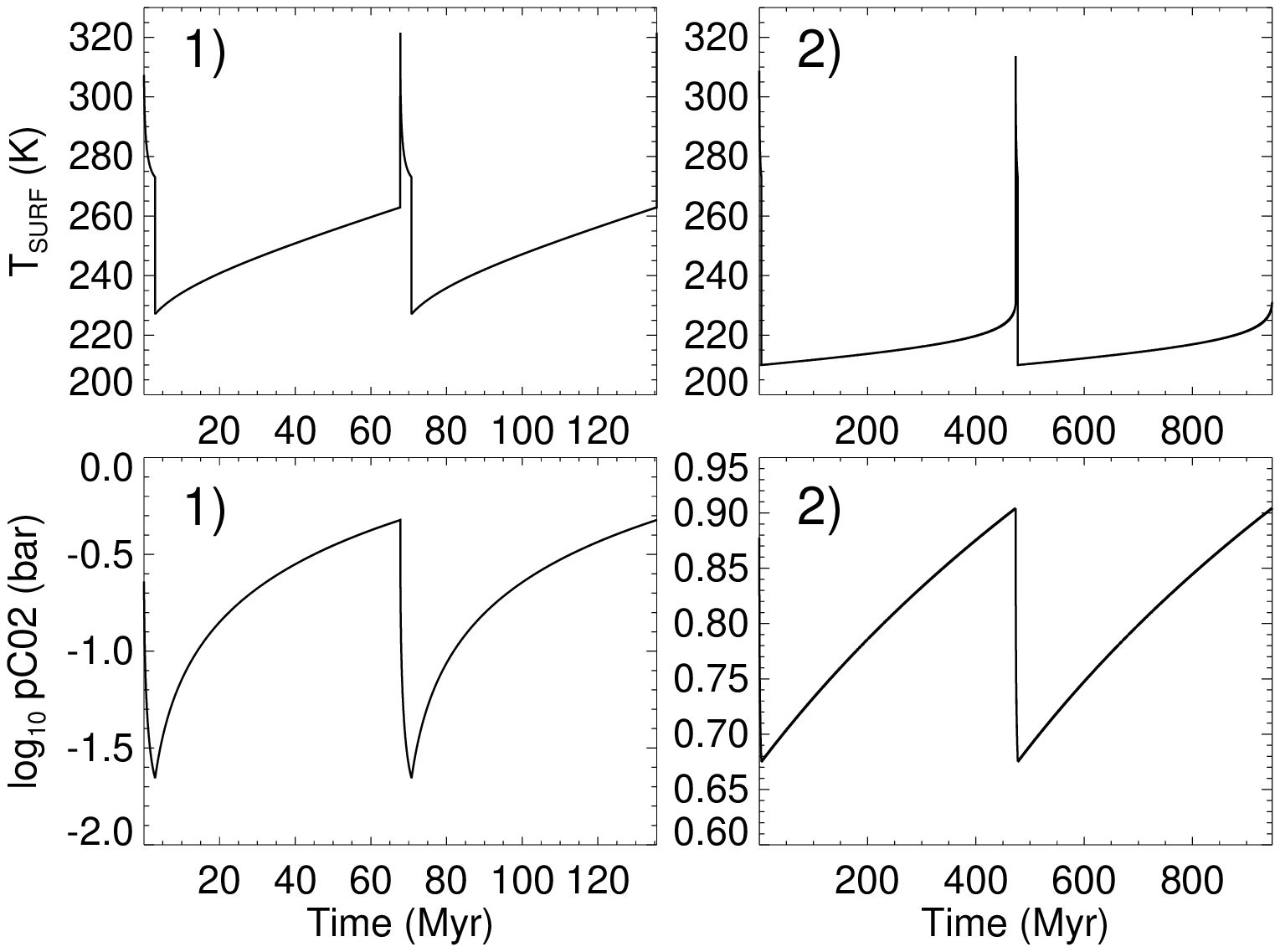}}
\centerline{\includegraphics[width=.55\textwidth]{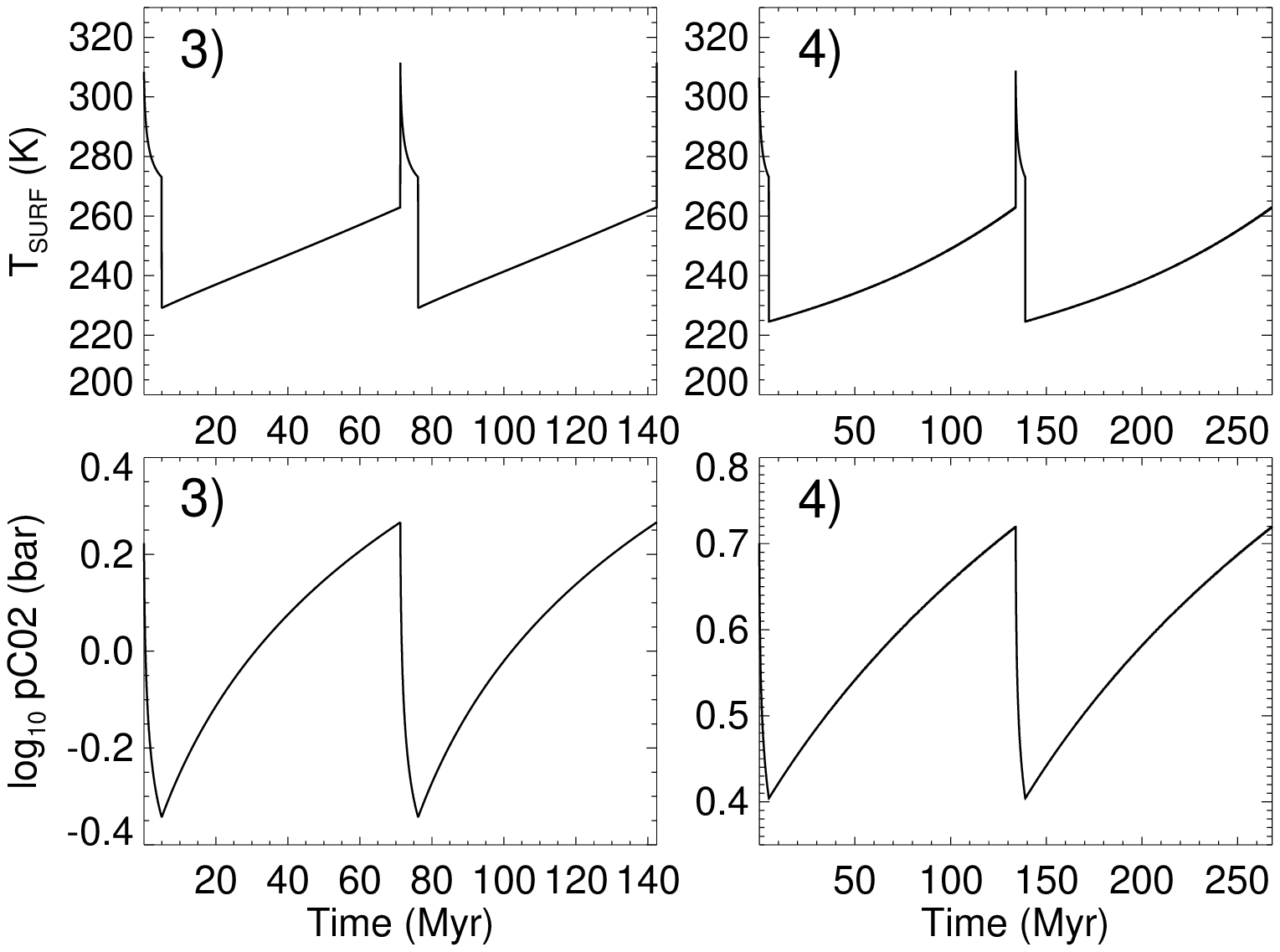}}
\centerline{\includegraphics[width=.55\textwidth]{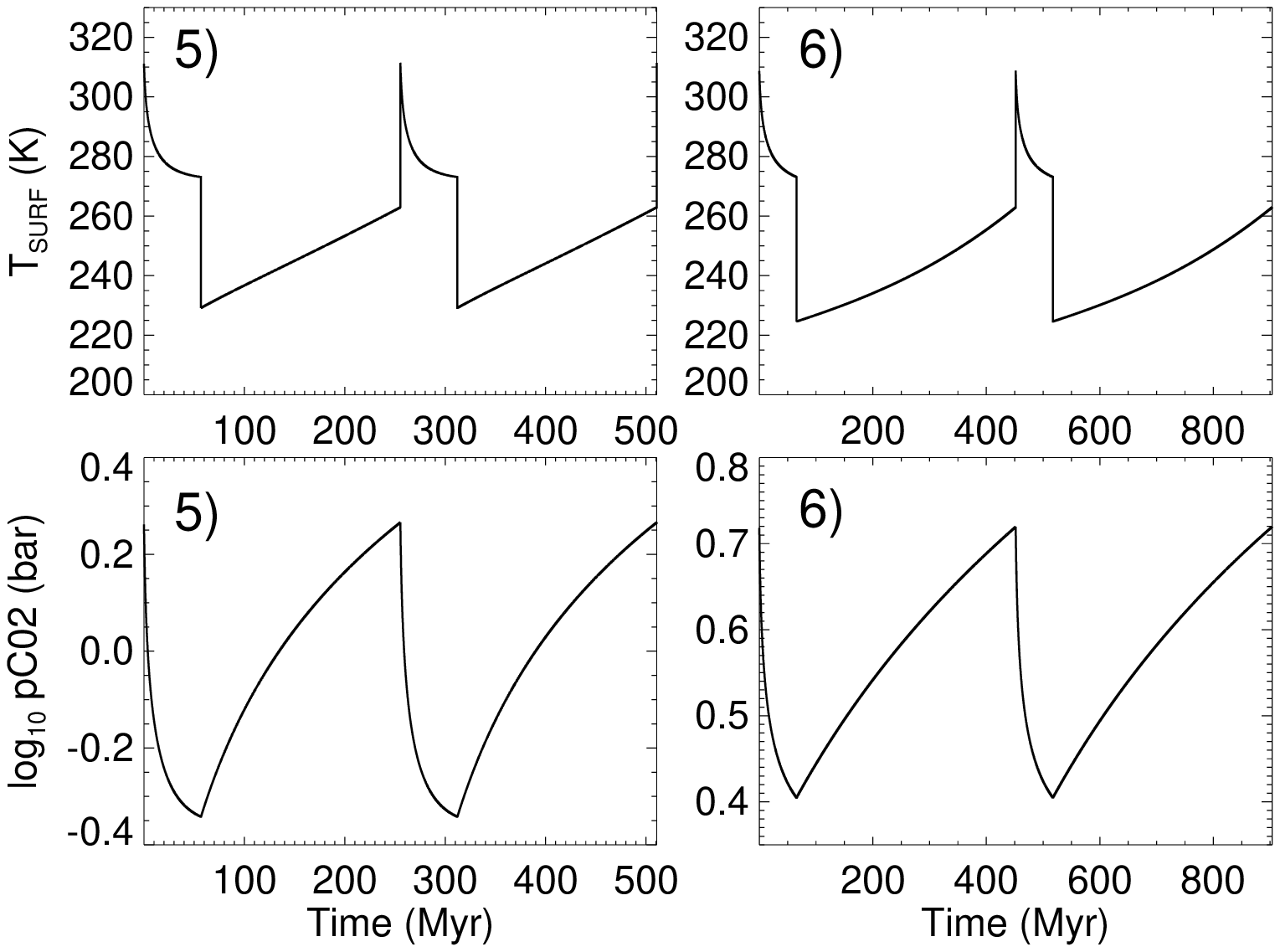}}
\caption{Six examples of time-evolved climate cycles, with two full
  cycles shown in each case. For each model (1-6), the evolution of
  atmospheric partial $CO_2$ pressure (lower panel) and surface
  temperature (upper panel) are shown. Model 1: Default weathering
  model at 1.1 AU ($\beta =0.5$ in Equation~(3)). Model 2: Default
  weathering model at 1.8 AU. Model 3: 3 times larger $CO_2$
  outgassing rate at 1.3 AU. Model 4: 3 times larger $CO_2$ outgassing
  rate at~1.6 AU. Model 5: Weaker $pCO_2$ weathering dependence
  ($\beta=0.25$) at 1.3 AU. Model 6: Weaker $pCO_2$ weathering
  dependence ($\beta=0.25$) at 1.6 AU.  Increasing the $CO_2$
  outgassing rate shortens the duration of the cold, $CO_2$ build-up
  phase. Weakening the weathering $pCO_2$ dependence ($\beta=0.25$
  rather than $0.5$) lengthens the duration of the hot weathering
  phase, resulting in a much larger fraction of time spent with
  surface liquid water.}\label{afoto3}
\end{center}
\end{figure}

\appendix

\section{Energy Balance Climate Model.}

We model the climate of Earth-like planets with a zero-dimensional
reduction of a one-dimensional energy balance model \cite{9}. The
model assumes Earth parameters unless otherwise specified (e.g.,
surface gravity, land/ocean fraction and nitrogen contribution to the
total atmospheric mass). Greenhouse effect from atmospheric $H_2O$ and
$CO_2$ are included, with an atmospheric vapor pressure set by surface
evaporation (temperature).

The top-of-atmosphere albedo and the outgoing longwave radiation flux
are modeled as polynomial fits to a large number of
radiative-convective models \cite{9}. The polynomial fits are
functions of surface temperature, partial $CO_2$ pressure, solar
zenith angle and surface albedo. Simple prescriptions for snow/ice
coverage, surface albedo and water cloud coverage are adopted
\cite{9}.  For simplicity, we fix the cosine of the zenith angle to
$\mu=0.4$ and the albedo of ice-free oceans to $0.07$ in all the
models presented here. We also smooth out the top-of-atmosphere albedo
polynomials near the $280$~K transition to improve the continuity of
the albedo function with temperature.

Based on published results \cite{2}, we expect that our results would
be quantitatively different, but remain qualitatively valid, for
planets that differ modestly from Earth in terms of their surface
gravity, land/ocean fraction and/or nitrogen atmospheric content. Note
that it has been suggested that weathering does not strongly depend on
land/ocean fraction on an Earth-like planet \cite{19}.

The energy balance model employed here may not be fully reliable
beyond $1.3$-$1.4$~AU, where $CO_2$ clouds are expected to form and
influence the climate \cite{9,10}. The most massive $CO_2$ atmospheres
found in our models only marginally approach hard limits on $CO_2$
condensation \cite{27}.

\section{Model Simplifications and Limitations.}

Our models are idealized in a number of important ways. In addition to
the simplified, zero-dimensional treatment of climate described above,
which suggests the possibility of richer behaviors in higher
complexity, three-dimensional climate models, our treatment of
weathering processes is intentionally simple, in order to isolate they
key factors that determine climate stability. We ignore seafloor weathering \cite{19} and the mantle $CO_2$ cycle \cite{28}.

The absolute calibration of weathering rates in the absence of land
vascular plants is unknown, but it is thought to be less than in their
presence \cite{11}. For concreteness, we have chosen to calibrate
weathering fluxes in our models using current Earth \cite{13,14}
(Equation~(3)). We note that in a model admitting steady-state
solutions, a factor three decrease in the weathering rate is
equivalent to a factor three increase in the volcanic outgassing rate
(see Equation~(2)), which is one of the cases shown in Figure~1
(dotted blue line). Such a model retains the main qualitative feature
highlighted in this work, which is the disappearance of stable
climates solutions at low enough insolation levels (beyond $1.2$AU for
the dotted blue line shown in Figure~1). Different calibrations in
weathering rates and/or volcanic outgassing rates will thus affect our
results quantitatively, but our main conclusions should remain valid.

More generally, a planet is likely to change its weathering regime
gradually over time, as different forms of life emerge and spread over
its surface \cite{3,12}. Our models have intentionally focused on the
distinction between the absence and presence of land vascular planets,
which exemplifies the interplay between life, weathering processes and
climate stability.




\end{document}